\documentclass[aps,prb,reprint,amsmath,amssymb,longbibliography,nofootinbib,floatfix]{revtex4-2}

\usepackage[T1]{fontenc}
\usepackage{lmodern}
\usepackage{bm}
\usepackage{graphicx}
\usepackage{booktabs}
\usepackage{siunitx}
\usepackage{microtype}
\usepackage{placeins}
\usepackage[hidelinks]{hyperref}
\graphicspath{{figures/}}

\newcommand{\kB}{k_{\mathrm B}}
\newcommand{\ii}{\mathrm i}
\newcommand{\dd}{\mathrm d}
\newcommand{\Tr}{\operatorname{Tr}}
\newcommand{\sech}{\operatorname{sech}}
\newcommand{\artanh}{\operatorname{artanh}}
\newcommand{\sgn}{\operatorname{sgn}}
\newcommand{\Om}{\Omega}
\newcommand{\mcM}{\mathcal M}

\begin{document}

\title{Evanescent-mode Casimir-Josephson force and gate-controlled resonances in ballistic graphene Josephson junctions}

\author{Shahrukh Salim}
\affiliation{Korea Institute for Advanced Study, Seoul, Republic of Korea}
\email{shahrukh@kias.re.kr}

\date{\today}

\begin{abstract}
We develop a microscopic scattering theory of the phase-dependent equilibrium mechanical response of a ballistic superconductor-graphene-superconductor Josephson junction. The calculation is based on an energy-dependent Dirac scattering matrix embedded in a normalized Matsubara determinant, so the reported force correction contains the complete Bogoliubov-de Gennes spectrum of the stated ideal model. At charge neutrality the full determinant approaches a closed-form evanescent-mode result proportional to $\Delta W/L^2$, reaching $2\ln 2\,\Delta W/(\pi L^2)$ at phase difference $\pi$ in the zero-temperature short-junction limit. Gate doping produces propagating channels and Fabry-P\'erot structure, causing large oscillations and sign reversals of the phase-dependent force correction $\delta F=F(\phi)-F(0)$. We distinguish the interband-to-intraband Andreev crossover, controlled by $|\mu|/\Delta$, from the evanescent-to-propagating crossover, controlled by $|\mu|L/(\hbar v_F)$. Exact real-energy subgap poles obtained from the same energy-dependent scattering matrix are used to diagnose specular/interband and retro/intraband character; these labels are not treated as separately measurable thermodynamic forces in the mixed regime. We also quantify the difference between the complete determinant result and the frozen-scattering short-junction approximation without identifying that difference with a pure continuum force. The resulting gate- and phase-dependent mechanical signal provides a Dirac-material extension of earlier superconductivity-induced mechanical-force proposals.
\end{abstract}

\maketitle

\section{Introduction}

A quantum spectrum that depends on a geometrical coordinate produces a thermodynamic force conjugate to that coordinate, the principle underlying the Casimir effect.\cite{Casimir1948,Lifshitz1956} In a superconducting weak link the quasiparticle spectrum depends both on the superconducting phase difference $\phi$ and on the junction length $L$. If $\Om(\phi,L)$ denotes the equilibrium grand potential, the Josephson current\cite{Josephson1962} and the longitudinal mechanical force satisfy
\begin{equation}
I(\phi,L)=\frac{2e}{\hbar}\left.\frac{\partial\Om}{\partial\phi}\right|_L,
\qquad
F(\phi,L)=-\left.\frac{\partial\Om}{\partial L}\right|_\phi.
\label{eq:thermo}
\end{equation}
Superconductivity-induced phase-controlled forces were predicted for ballistic Josephson nanowires by Krive \textit{et al.}\cite{Krive2004} More recently, the Casimir-Josephson force of a short metallic point contact was formulated in scattering language and shown to receive an essential, and in that model dominant, contribution from above-gap quasiparticles.\cite{BeenakkerCJ2023,Levchenko2025} These results make clear that a mechanical force probes spectral length sensitivity rather than merely the states that dominate the supercurrent.

Graphene provides a qualitatively different weak link because its low-energy quasiparticles carry conduction/valence-band and sublattice-pseudospin structure.\cite{Novoselov2005,CastroNeto2009} At a graphene-superconductor interface, the Andreev electron-to-hole conversion\cite{Andreev1964} can be intraband (retro Andreev reflection) or interband (specular Andreev reflection).\cite{BeenakkerSAR2006,BeenakkerRMP2008,Efetov2016} The ballistic SGS spectrum and current-phase relation are well established,\cite{Titov2006,BlackSchaffer2008} and our earlier work analyzed the Andreev spectrum and transport response across retro and specular regimes in SGS junctions.\cite{Salim2023} Ballistic graphene Josephson physics has also been observed experimentally in gate-tunable devices.\cite{Heersche2007,Calado2015,Borzenets2016,Nanda2017,Bretheau2017}

The mechanical problem introduces an additional derivative with respect to $L$. Near charge neutrality, the transmission eigenvalues of a wide graphene strip are evanescent, $\tau(q)=\sech^2(qL)$, and therefore retain an explicit length dependence even in the nominal short-junction limit. At larger carrier density, propagating channels acquire the Fabry-P\'erot phase $k_xL$. These two mechanisms suggest a force response not obtained by simply transplanting a length-independent metallic point-contact result to graphene. Moreover, the Andreev band-character crossover and the propagation crossover are governed by different dimensionless variables: $|\mu|/\Delta$ and $|\mu|L/(\hbar v_F)$, respectively.

A second issue is methodological. The familiar short-junction expression $E_n=\Delta\sqrt{1-\tau_n\sin^2(\phi/2)}$ freezes the normal-region scattering matrix at the Fermi energy. Subtracting the force obtained from that approximation from a full energy-dependent calculation does \emph{not}, in general, isolate a pure continuum force: the difference also contains finite-$L/\xi$ corrections to subgap levels. We therefore use the complete Matsubara determinant as the primary thermodynamic quantity and refer to its difference from the frozen-scattering result only as a \emph{beyond-short-junction remainder}. For spectral interpretation we additionally solve the exact real-energy secular equation for subgap poles. This choice avoids assigning a physically misleading continuum label to a mixed correction.

\textit{Relation to prior work.} Earlier studies established phase-controlled superconductivity-induced mechanical forces and the importance of energy-dependent above-gap quasiparticles in metallic weak links,\cite{Krive2004,BeenakkerCJ2023,Levchenko2025} while graphene studies established specular and retro Andreev conversion together with the SGS Andreev spectrum and Josephson response.\cite{BeenakkerSAR2006,Titov2006,Salim2023} The present work connects these strands by calculating the phase-dependent mechanical response of a ballistic Dirac weak link with a fully energy-dependent scattering determinant. The graphene-specific results are the neutral evanescent-mode law proportional to $\Delta W/L^2$, the gate-controlled onset of coherent propagation resonances, and the separation of the band-character scale $|\mu|/\Delta$ from the propagation scale $|\mu|L/(\hbar v_F)$. Exact real-energy subgap poles are used to verify the microscopic Andreev character without treating interband and intraband sectors as separately measurable forces.

Here we calculate the phase-dependent force correction $\delta F(\phi,L)=F(\phi,L)-F(0,L)$ of an ideal ballistic monolayer SGS junction with heavily doped $s$-wave contacts. At charge neutrality the full determinant approaches the analytical evanescent-mode law $\delta F\propto\Delta W/L^2$. Gate doping produces coherent resonances and repeated sign changes of $\delta F$, while exact subgap poles expose the crossover from interband/specular to intraband/retro character near $|\mu|\sim\Delta$. The propagation resonances occur on the separate scale $|\mu|L/(\hbar v_F)\sim1$. Recent work showing that the electrostatic profile can materially alter graphene Josephson transport\cite{Rycerz2026} motivates our explicit discussion of which ideal-interface predictions are robust and which are expected to broaden in realistic devices. A recent graphene-based study of specular Andreev reflection in a different superconducting platform further illustrates the continuing relevance of microscopic interband diagnostics.\cite{Li2025}

The term \emph{complete} below means complete with respect to the BdG quasiparticle spectrum of the stated single-particle scattering model. It does not include the ordinary electromagnetic Casimir force, electrostatic backgrounds, elastic stresses, electron-electron interactions, self-consistent order-parameter suppression, disorder, or phonons.

\section{Microscopic model}

\subsection{Dirac-Bogoliubov-de Gennes description}

\begin{figure}[!tb]
\centering
\includegraphics[width=\linewidth]{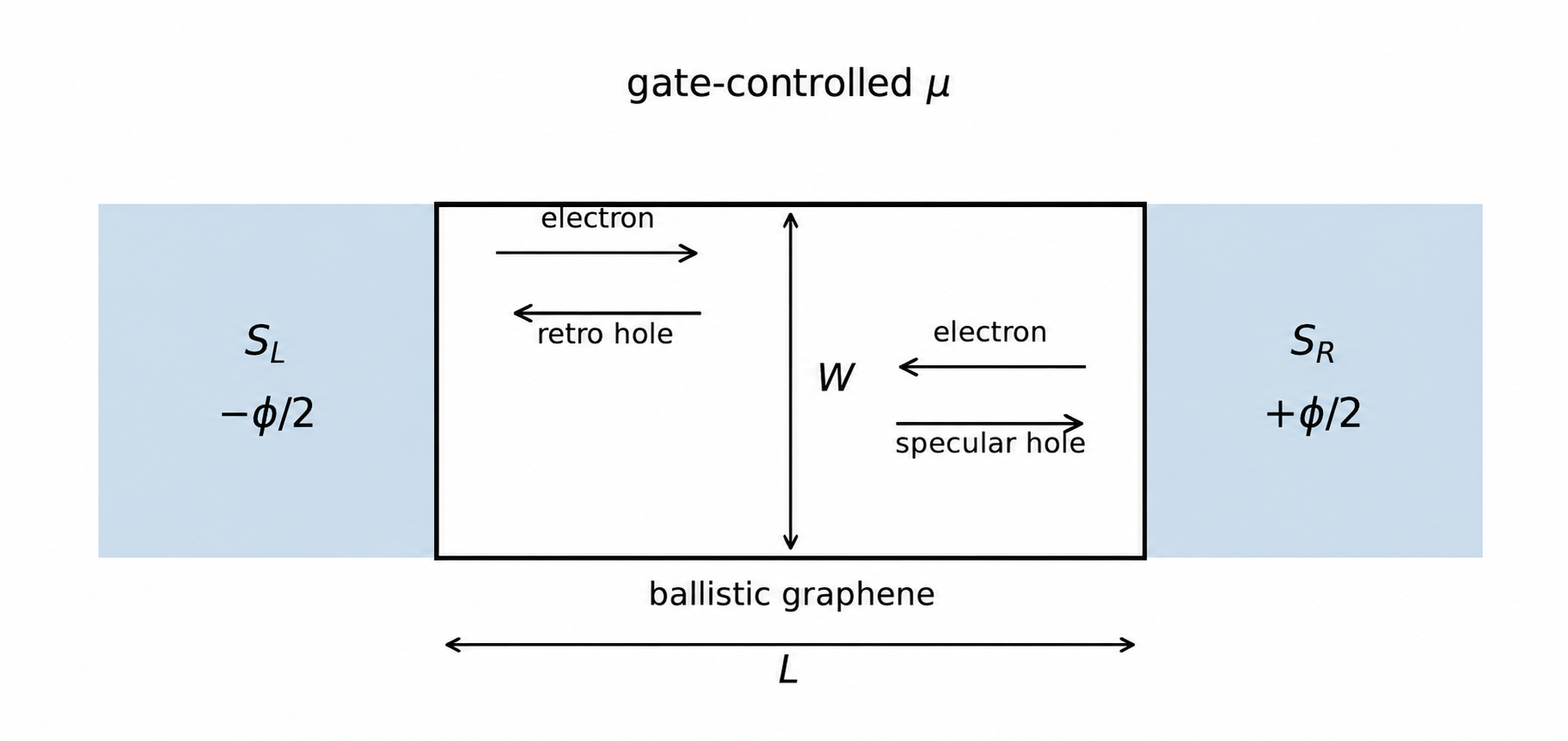}
\caption{Geometry of the ballistic superconductor-graphene-superconductor (SGS) junction. A ballistic graphene strip of length $L$ and width $W$ is contacted by two heavily doped $s$-wave superconductors held at phases $-\phi/2$ and $+\phi/2$, and a gate tunes the graphene chemical potential $\mu$. At each interface an incident electron is Andreev reflected as a hole: retro reflection (left) stays within the same Dirac band, whereas specular reflection (right) converts to the opposite band. The arrows indicate band character only; they are not semiclassical trajectories, which do not exist for the evanescent modes that dominate at charge neutrality.}
\label{fig:geometry}
\end{figure}

We consider a monolayer graphene strip of length $L$ and width $W$ between two conventional superconducting electrodes [Fig.~\ref{fig:geometry}]. The pair potential is modeled as
\begin{equation*}
\Delta(x)=
\begin{cases}
\Delta e^{-\ii\phi/2}, & x<0,\\
0, & 0<x<L,\\
\Delta e^{+\ii\phi/2}, & x>L.
\end{cases}
\end{equation*}
The normal graphene region is ballistic and described near one valley by
\begin{equation*}
h_G=\hbar v_F(\sigma_x k_x+\sigma_y k_y)-\mu,
\end{equation*}
where $\mu$ is measured from the Dirac point. The Dirac-Bogoliubov-de Gennes equation is
\begin{equation}
\begin{pmatrix}
h_G & \Delta(x)\\
\Delta^*(x) & -h_G^*
\end{pmatrix}
\begin{pmatrix}u\\v\end{pmatrix}
=E\begin{pmatrix}u\\v\end{pmatrix}.
\label{eq:dbdg}
\end{equation}
Spin and valley give the degeneracy factor $g=4$. Infinite-mass transverse confinement is used for definiteness,
\begin{equation*}
q_n=\frac{\pi}{W}\left(n+\frac12\right),\qquad n=0,1,2,\ldots.
\end{equation*}
In the wide-junction limit the leading results do not depend on this specific edge convention.

The natural coherence length and dimensionless control parameters are
\begin{equation}
\xi=\frac{\hbar v_F}{\Delta},\qquad
\frac{L}{\xi},\quad \frac{W}{L},\quad
\frac{\mu L}{\hbar v_F},\quad
\frac{\mu}{\Delta},\quad
\frac{\kB T}{\Delta}.
\label{eq:parameters}
\end{equation}
The two gate variables in Eq.~\eqref{eq:parameters} play different roles: $\mu/\Delta$ controls the band character of Andreev conversion, whereas $\mu L/(\hbar v_F)$ controls longitudinal propagation and interference.

\subsection{Energy-dependent graphene scattering matrix}

For each transverse mode $q$, the scattering matrix of a graphene strip connected to heavily doped normal leads can be written at complex excitation energy $z$ as
\begin{equation}
S_e(z;q)=
\begin{pmatrix}
r_e&t_e\\ t_e&r_e
\end{pmatrix},
\label{eq:Se}
\end{equation}
with
\begin{align}
k_e(z)&=\frac{\mu+z}{\hbar v_F},
\nonumber\\
k_{xe}(z)&=\sqrt{k_e^2-q^2},
\nonumber\\
t_e(z)&=\frac{k_{xe}}
{k_{xe}\cos(k_{xe}L)-\ii k_e\sin(k_{xe}L)},
\label{eq:t}\\
r_e(z)&=\frac{-\ii q\sin(k_{xe}L)}
{k_{xe}\cos(k_{xe}L)-\ii k_e\sin(k_{xe}L)}.
\label{eq:r}
\end{align}
The square-root branch is chosen with $\operatorname{Im}k_{xe}\geq0$, ensuring decaying rather than growing solutions after analytic continuation. For real energy and a propagating mode, Eqs.~\eqref{eq:t} and \eqref{eq:r} give a unitary matrix. At zero energy the transmission probability is
\begin{equation}
\begin{split}
\tau(q)&=\left[1+\frac{q^2}{k_x^2}\sin^2(k_xL)\right]^{-1},\\
k_x&=\sqrt{\left(\frac{\mu}{\hbar v_F}\right)^2-q^2}.
\end{split}
\label{eq:tau}
\end{equation}
For $q>|\mu|/(\hbar v_F)$ one sets $k_x=\ii\kappa$, converting the trigonometric functions into hyperbolic functions. At charge neutrality,
\begin{equation}
\tau(q)=\sech^2(qL).
\label{eq:neutral_tau}
\end{equation}
In the absence of magnetic field the hole scattering matrix on the imaginary axis follows from particle-hole symmetry,
\begin{equation*}
S_h(\ii\omega;q)=S_e^*(-\ii\omega;q).
\end{equation*}

\section{Matsubara free energy and force}

\subsection{Normalized scattering determinant}

Define the phase matrix
\begin{equation*}
R_\phi=\begin{pmatrix}e^{\ii\phi/2}&0\\0&e^{-\ii\phi/2}\end{pmatrix}
\end{equation*}
and the positive real Matsubara Andreev amplitude
\begin{equation*}
a_m=\frac{\Delta}{\omega_m+\sqrt{\omega_m^2+\Delta^2}},
\qquad
\omega_m=(2m+1)\pi\kB T.
\end{equation*}
The factor $-\ii$ in the analytically continued Andreev amplitude produces the plus sign in
\begin{equation}
\mcM_{n,m}(\phi)=
1+a_m^2 R_\phi^\dagger S_h(\ii\omega_m;q_n)
R_\phi S_e(\ii\omega_m;q_n).
\label{eq:M}
\end{equation}
The phase-dependent grand potential, normalized to vanish at $\phi=0$, is
\begin{equation}
\boxed{
\delta\Om(\phi,L)=
-g\kB T\sum_{n=0}^{\infty}\sum_{m=0}^{\infty}
\ln\frac{\det\mcM_{n,m}(\phi)}{\det\mcM_{n,m}(0)}.
}
\label{eq:OmegaM}
\end{equation}
Equation~\eqref{eq:OmegaM} is the scattering-determinant form of the equilibrium grand potential of a phase-biased weak link;\cite{Beenakker1991,FurusakiTsukada1991} it includes both poles associated with Andreev bound states and the phase shift of continuum quasiparticles. The phase subtraction removes the dominant phase-independent ultraviolet contribution. The sign convention and the normalization are fixed by requiring that Eq.~\eqref{eq:OmegaM} reduce to the universal short-junction spectrum when the normal scattering matrix is energy independent, as shown in Appendix~\ref{app:reduction}.

The observable studied here is
\begin{equation}
\delta F(\phi,L)=
-\left.\frac{\partial\delta\Om}{\partial L}\right|_{\phi,\mu,W,\Delta,T}.
\label{eq:Fdef}
\end{equation}
The derivative is taken at fixed physical width and chemical potential. Differentiating the trace-log gives
\begin{align}
\delta F(\phi,L)=g\kB T\sum_{n,m}\operatorname{Re}\Big\{&
\Tr[\mcM^{-1}(\phi)\partial_L\mcM(\phi)]
\nonumber\\
&-\Tr[\mcM^{-1}(0)\partial_L\mcM(0)]\Big\}.
\label{eq:Ftrace}
\end{align}
In the numerical implementation we evaluate Eq.~\eqref{eq:OmegaM} at $L\pm\delta L$ with the same transverse-mode set and use a centered derivative. Convergence is established by reducing $\delta L$ and independently enlarging the transverse-mode and Matsubara cutoffs. Equation~\eqref{eq:Ftrace} is retained as the analytical trace-log identity underlying that derivative rather than being presented as a separately implemented numerical check.

The current obtained from the same determinant is
\begin{equation*}
I(\phi,L)=\frac{2e}{\hbar}\frac{\partial\delta\Om}{\partial\phi}.
\end{equation*}
Consequently,
\begin{equation*}
\frac{\partial\delta F}{\partial\phi}
=-\frac{\hbar}{2e}\frac{\partial I}{\partial L},
\end{equation*}
which provides an additional thermodynamic identity that can be used for future independent code validation.

\subsection{Short-junction benchmark, exact poles, and the remainder}

If the normal scattering matrix is frozen at the Fermi energy, the positive short-junction Andreev level of a transmission channel is
\begin{equation}
E_n^{(0)}(\phi)=\Delta\sqrt{1-\tau_n\sin^2\frac{\phi}{2}}.
\label{eq:ABS}
\end{equation}
Its phase-dependent free energy is
\begin{equation*}
\delta\Om_{\mathrm{ABS}}^{(0)}=-g\kB T\sum_n
\ln\frac{\cosh[E_n^{(0)}(\phi)/(2\kB T)]}
{\cosh[\Delta/(2\kB T)]}.
\end{equation*}
We use the corresponding force $\delta F_{\mathrm{ABS}}^{(0)}$ as a controlled short-junction benchmark. The quantity
\begin{equation}
\delta F_{\mathrm{rem}}=\delta F_{\mathrm{full}}-\delta F_{\mathrm{ABS}}^{(0)}
\label{eq:remainder}
\end{equation}
is called the \emph{beyond-short-junction remainder}. It contains all corrections generated by the energy dependence of the normal scattering matrix, including both corrections to the subgap spectrum and above-gap phase shifts, and is not identified with a pure continuum force.

For microscopic information about the actual subgap spectrum we analytically continue the same scattering problem to real $|E|<\Delta$. In the same zero-field particle-hole-symmetric model, the real-energy hole matrix obeys
\begin{equation*}
S_h(E;q)=S_e^*(-E;q).
\end{equation*}
With $\alpha(E)=\exp[-\ii\arccos(E/\Delta)]$, define the round-trip matrix
\begin{equation*}
\mathcal Q_n(E,\phi)=R_\phi^\dagger S_h(E;q_n)R_\phi S_e(E;q_n).
\end{equation*}
The exact positive subgap poles satisfy
\begin{equation}
D_n(E,\phi)\equiv\det\left[1-\alpha^2(E)\mathcal Q_n(E,\phi)\right]=0.
\label{eq:secular}
\end{equation}
Equation~\eqref{eq:secular} uses the full energy-dependent $S_e$ and $S_h$ rather than $S(E=0)$. Numerically, poles are located on a grid uniform in the Andreev angle $\arccos(E/\Delta)$ and refined by bounded minimization of $|D_n|$; zero multiplicities are checked from the singular values of the secular matrix. This pole solver is used below for the interband/intraband diagnostic and to verify that the real-energy secular structure is consistent with the same energy-dependent scattering matrices used on the Matsubara axis.

\section{Interband and intraband Andreev character}

For a positive exact quasiparticle energy $E$, the electron and hole band indices are
\begin{equation*}
s_e=\sgn(\mu+E),\qquad s_h=\sgn(\mu-E).
\end{equation*}
Thus $s_es_h=+1$ corresponds to intraband/retro conversion and $s_es_h=-1$ to interband/specular conversion. For electron-doped graphene this reduces to $E<|\mu|$ for retro and $E>|\mu|$ for specular conversion. At exact neutrality every finite-energy conversion is interband, although the finite-junction states that dominate the force are evanescent and do not possess a literal semiclassical reflection trajectory.

A unique additive thermodynamic decomposition of the \emph{total} force into retro and specular pieces does not generally exist in the crossover regime because the determinant combines coherent amplitudes. Accordingly, Fig.~\ref{fig:character} labels the exact poles of Eq.~\eqref{eq:secular} by their band character rather than constructing phenomenological partial forces. This is a microscopic spectral diagnostic, while all force plots use the complete determinant of Eq.~\eqref{eq:OmegaM}.

\section{Analytical charge-neutral benchmark}

At $\mu=0$ and $T=0$, Eqs.~\eqref{eq:neutral_tau} and \eqref{eq:ABS} give
\begin{equation*}
E_q(\phi)=\Delta\sqrt{1-\sech^2(qL)\sin^2\frac{\phi}{2}}.
\end{equation*}
In the wide-junction limit,
\begin{equation*}
\sum_n\longrightarrow \frac{W}{\pi}\int_0^\infty\dd q.
\end{equation*}
Subtracting the $\phi=0$ energy and using the positive-energy BdG convention gives
\begin{align}
\delta\Om_{\mathrm{ABS}}(\phi,L)
&=\frac{2\Delta W}{\pi L}\,\mathcal G(\phi),
\label{eq:neutralOmega}\\
\mathcal G(\phi)
&=s\artanh s+\frac12\ln(1-s^2),
\label{eq:G}\\
s&=\left|\sin\frac{\phi}{2}\right|.\nonumber
\end{align}
The force is therefore
\begin{equation}
\boxed{
\delta F_{\mathrm{ABS}}(\phi,L)=
\frac{2\Delta W}{\pi L^2}\,\mathcal G(\phi).
}
\label{eq:neutralF}
\end{equation}
At $\phi=\pi$, the separately divergent terms in Eq.~\eqref{eq:G} combine to give
\begin{equation*}
\delta F_{\mathrm{ABS}}(\pi,L)=
\frac{2\ln2}{\pi}\frac{\Delta W}{L^2}.
\end{equation*}
For small phase,
\begin{equation*}
\mathcal G(\phi)=\frac{\phi^2}{8}+\mathcal O(\phi^4),
\end{equation*}
whereas $\mathcal G(\phi)$ approaches $\ln2$ nonanalytically as $\phi\rightarrow\pi$. Finite temperature rounds this endpoint behavior.

The physical reason for Eq.~\eqref{eq:neutralF} is the explicit length derivative
\begin{equation}
\partial_L\tau(q)=-2q\sech^2(qL)\tanh(qL).
\label{eq:dtauneutral}
\end{equation}
In an ordinary ideal point contact, the channel transparency is often modeled as independent of $L$, eliminating this leading frozen-scattering bound-state force and making energy-dependent corrections, including above-gap quasiparticles, essential. Neutral graphene differs because the evanescent transmission itself changes on the scale $L$, so the complete short-junction force already approaches the evanescent-mode expression of Eq.~\eqref{eq:neutralF}. We do not infer a unique bound-versus-continuum hierarchy from this observation alone.

\section{Numerical results}

All figures use $\Delta=1$ and $\xi=\hbar v_F/\Delta=1$ internally. Unless stated otherwise, $W/L=20$, $\kB T/\Delta=0.05$, ideal interfaces, and heavily doped superconducting contact regions are assumed. The force is normalized by $\Delta W/L^2$. Parameters are summarized in Table~\ref{tab:parameters}.

\begin{table*}[t]
\caption{Parameters used in the numerical figures. For every mechanical derivative, $W$, $\mu$, $T$, and $\Delta$ are held fixed while $L$ is varied. In a length-scaling plot, each point represents a different device with $W/L=20$.}
\label{tab:parameters}
\begin{ruledtabular}
\begin{tabular}{lccccc}
Figure & $L/\xi$ & $W/L$ & $\kB T/\Delta$ & gate range & purpose\\
\hline
2 & 0.20 & 20 & 0.05 & $\mu=0$ & neutral analytical validation\\
3 & 0.50 & 20 & 0.05 & $0\leq\mu L/\hbar v_F\leq6$ & gate-dependent force correction\\
4 & 0.50 & 20 & 0.05 & $0\leq\mu L/\hbar v_F\leq6$ & phase-gate force landscape\\
5 & $0.05$ to $1.2$ & 20 & 0.05 & $\mu=0$ & neutral length scaling\\
6 & $0.12$ to $0.75$ & 20 & 0.05 & $\mu=0$ & beyond-short-junction remainder\\
7 & 0.30 & 20 & $0.03$ to $0.45$ & $\mu=0$ & thermal suppression\\
8 & 0.30 & 20 & 0.05 & $0\leq|\mu|/\Delta\leq1.4$ & force-weighted interband diagnostic\\
9 & 0.30 & 20 & 0.05 & $0\leq|\mu|/\Delta\leq1.4$ & exact real-energy Andreev character\\
\end{tabular}
\end{ruledtabular}
\end{table*}

\subsection{Validation at charge neutrality}

\begin{figure}[!tb]
\centering
\includegraphics[width=\linewidth]{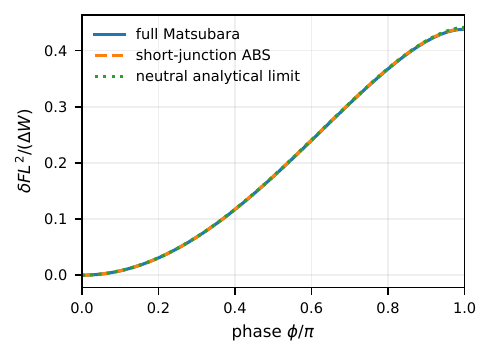}
\caption{Validation at charge neutrality. Normalized force $\delta F\,L^2/(\Delta W)$ versus phase $\phi/\pi$ for $L/\xi=0.20$, $W/L=20$, and $\kB T/\Delta=0.05$. The full Matsubara determinant (solid), which includes both discrete Andreev bound states and continuum quasiparticles, agrees closely with the discrete-ABS sum (dashed) and with the zero-temperature analytical evanescent-mode law of Eq.~\eqref{eq:neutralF} (dotted). The small residual difference between the full and frozen-scattering curves is the beyond-short-junction remainder. The agreement establishes that the leading neutral short-junction force is captured by the evanescent transmission spectrum, without requiring a separate bound/continuum force assignment.}
\label{fig:neutral}
\end{figure}

Figure~\ref{fig:neutral} compares three calculations: the full Matsubara determinant, the frozen-scattering spectrum of Eq.~\eqref{eq:ABS}, and the zero-temperature analytical expression in Eq.~\eqref{eq:neutralF}. At $L/\xi=0.20$ the curves are almost indistinguishable over the full phase interval. At $\phi=\pi$ the normalized forces are
\begin{align*}
\frac{\delta F L^2}{\Delta W}
&=0.43798\quad\text{(full)},\\
&=0.43876\quad\text{(ABS)},\\
&=0.44127\quad\text{($T=0$ analytical)}.
\end{align*}
The small difference between the two ABS values is thermal; the residual difference between the full determinant and the frozen-scattering ABS result is the beyond-short-junction remainder of Eq.~\eqref{eq:remainder}. This comparison verifies the determinant normalization, degeneracy convention, imaginary-energy branch choice, and mechanical sign convention without assigning that small remainder to a unique continuum sector.

As an independent phase-derivative benchmark, we also evaluate the Josephson current from $I=(2e/\hbar)\partial_\phi\delta\Omega$ at charge neutrality and compare it with the Titov--Beenakker short-junction limit,\cite{Titov2006}
\begin{equation*}
I_{\rm TB}(\phi)=\frac{2e\Delta W}{\pi\hbar L}\cos\frac{\phi}{2}\,\operatorname{artanh}\!\left(\sin\frac{\phi}{2}\right).
\end{equation*}
For $L/\xi=0.02$, $W/L=20$, and $k_{\rm B}T/\Delta=0.01$, the numerical current at $\phi/\pi=0.2,0.4,0.6,$ and $0.8$ differs from the zero-temperature limiting expression by less than $0.8\%$. This provides an independent check of the phase convention and spin-valley degeneracy; the remaining difference is consistent with the finite-$L/\xi$ and finite-temperature corrections retained by the determinant.

\subsection{Gate-induced oscillations and energy-dependent corrections}

\begin{figure}[!tb]
\centering
\includegraphics[width=\linewidth]{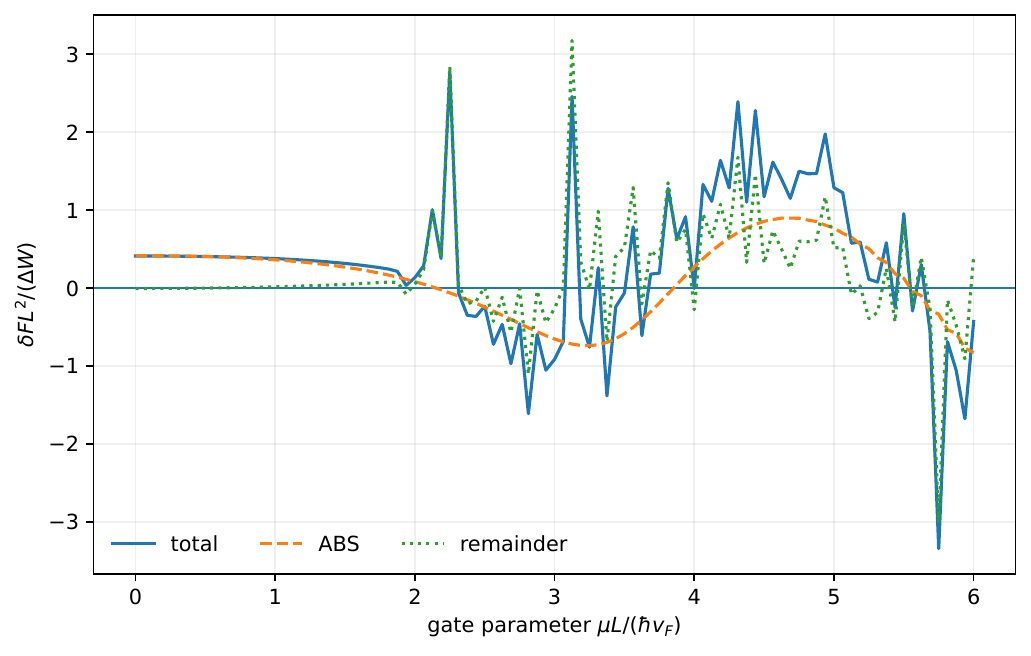}
\caption{Gate dependence of the phase-dependent force correction at $L/\xi=0.50$, $W/L=20$, $\phi=0.9\pi$, and $\kB T/\Delta=0.05$. Solid: full Matsubara force; dashed: frozen-scattering ABS benchmark; dotted: beyond-short-junction remainder. The remainder is not identified with a pure continuum force, and the sign shown is that of $\delta F=F(\phi)-F(0)$.}
\label{fig:gate}
\end{figure}

Figure~\ref{fig:gate} shows $\delta F$ at $\phi=0.9\pi$ versus $\mu L/(\hbar v_F)$ for $L/\xi=0.50$. Near neutrality the full response is smooth and close to the frozen-scattering short-junction benchmark. Once propagating channels are established, the phase $k_xL$ in the normal scattering matrix produces coherent resonances. The length derivative converts these resonances into alternating positive and negative peaks of the \emph{phase-dependent correction} $\delta F$.

The dotted curve in Fig.~\ref{fig:gate} is $\delta F_{\mathrm{rem}}$ from Eq.~\eqref{eq:remainder}. It is deliberately not called a continuum force. Around $\mu L/(\hbar v_F)=2.25$, the normalized full response is approximately $2.8$ while the frozen-scattering ABS benchmark is close to zero, showing that energy-dependent scattering is essential at that resonance. At $\mu L/(\hbar v_F)=3$, the normalized full result is approximately $-0.91$ compared with a short-junction value near $-0.65$. Thus gate voltage can reverse the sign of $\delta F$ at fixed phase, but this statement does not imply reversal of the absolute phase-independent mechanical force.

The sharpest structures are specific to the ideal coherent strip with abrupt, perfectly transmitting contacts. Smooth electrostatic profiles, finite mean free path, contact inhomogeneity, and gate averaging are expected to broaden them.\cite{Rycerz2026} The robust statement is the strong gate sensitivity of the complete force and the appearance of sign-changing phase-dependent resonances once propagating modes are available.

\FloatBarrier
\subsection{Phase-gate force landscape}

\begin{figure}[!tb]
\centering
\includegraphics[width=\linewidth]{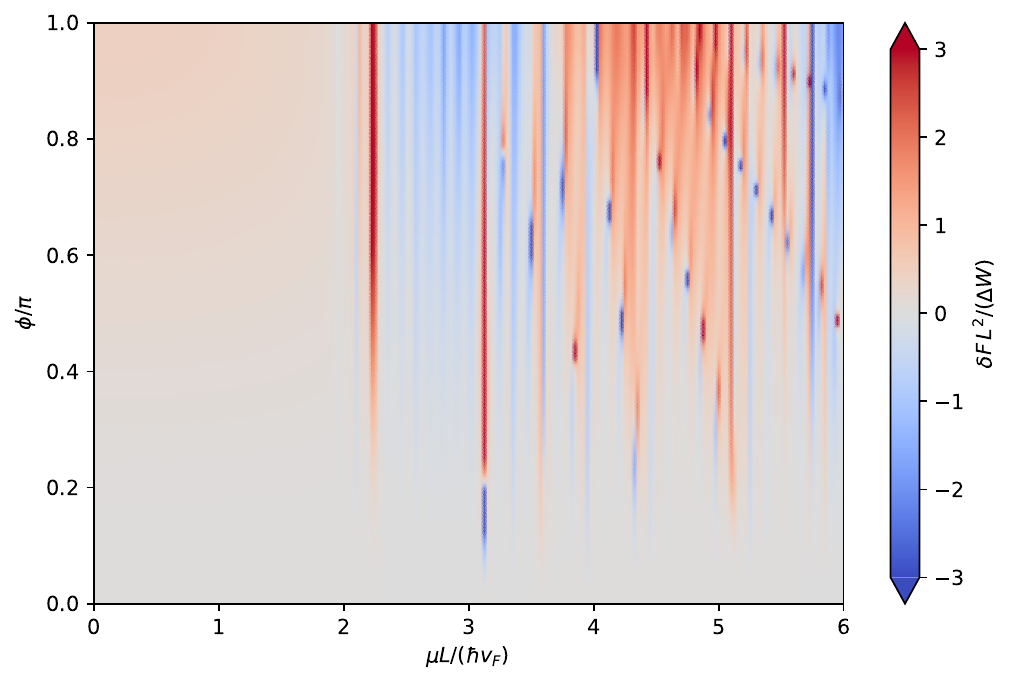}
\caption{Phase-gate landscape of $\delta F L^2/(\Delta W)$ for $L/\xi=0.50$ and $W/L=20$. The uncluttered color map emphasizes the smooth neutral regime and the narrow gate-induced Fabry--P\'erot structures at larger doping. Red and blue regions denote positive and negative values of the phase-dependent correction $\delta F=F(\phi)-F(0)$; the sign does not refer to the absolute force. The color scale is clipped at $\pm3$ to retain contrast in the background while the sharpest resonances extend beyond this range.}
\label{fig:landscape}
\end{figure}

The full phase-gate landscape is shown in Fig.~\ref{fig:landscape}. The force vanishes at $\phi=0$ by construction. Close to charge neutrality it grows monotonically with phase and remains positive under the convention of Eq.~\eqref{eq:Fdef}. At larger gate doping $\delta F$ changes sign along a proliferating set of zero-correction loci. These sign changes provide a direct experimental target: sweeping gate voltage at fixed phase reverses the phase-locked \emph{modulated} signal while leaving a static phase-independent background unchanged.

Electron-hole symmetry of the ideal Dirac model implies
\begin{equation*}
\delta F(\mu,\phi)=\delta F(-\mu,\phi),
\end{equation*}
which holds only for the symmetric Dirac model and is satisfied numerically to machine precision. A measurable asymmetry would therefore diagnose contact asymmetry, next-nearest-neighbor hopping, unequal electron and hole doping profiles, or interaction effects beyond the model.

\FloatBarrier
\subsection{Length scaling and the short-junction remainder}

\begin{figure}[!tb]
\centering
\includegraphics[width=\linewidth]{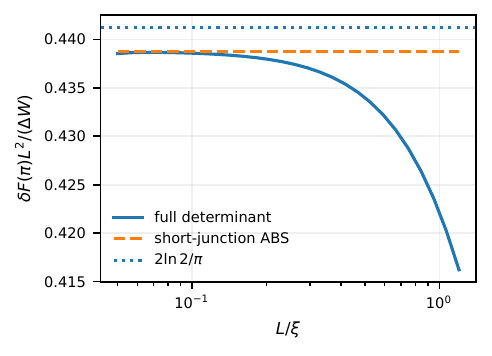}
\caption{Neutral length scaling at $\phi=\pi$, $W/L=20$, and $\kB T/\Delta=0.05$. The full Matsubara result and frozen-scattering benchmark both approach $2\ln2/\pi$ as $L/\xi\to0$, establishing the leading $\Delta W/L^2$ evanescent-mode scale.}
\label{fig:length}
\end{figure}

Figure~\ref{fig:length} tests the neutral short-junction scaling. For each plotted device $W/L=20$, while the derivative defining the force is taken at fixed physical $W$. As $L/\xi\rightarrow0$, the full result approaches the finite-temperature short-junction benchmark and the zero-temperature constant $2\ln2/\pi$. This establishes the leading $\Delta W/L^2$ law directly at the level of the complete Matsubara determinant.

\begin{figure}[!tb]
\centering
\includegraphics[width=\linewidth]{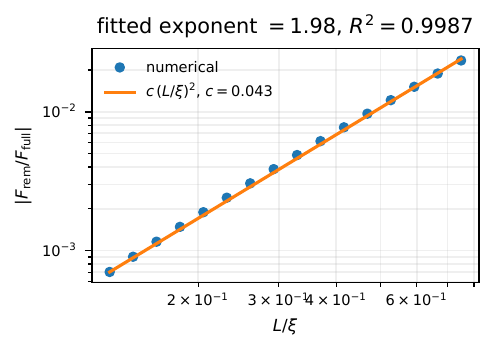}
\caption{Beyond-short-junction remainder at charge neutrality and $\phi=\pi$. The fit over $0.12\le L/\xi\le0.75$ gives $|F_{\rm rem}/F_{\rm full}|\simeq0.043(L/\xi)^{1.98}$. This is an approximation remainder, not a separately evaluated continuum force.}
\label{fig:remainder}
\end{figure}

The difference between the full and frozen-scattering calculations is quantified in Fig.~\ref{fig:remainder}. Over the numerical window $0.12\leq L/\xi\leq0.75$, the magnitude of the normalized remainder is accurately fit by
\begin{equation}
\left|\frac{F_{\mathrm{rem}}}{F_{\mathrm{full}}}\right|
\simeq c\left(\frac{L}{\xi}\right)^p,
\qquad c\simeq0.043,\quad p\simeq1.98.
\label{eq:remfit}
\end{equation}
Equation~\eqref{eq:remfit} is an empirical statement about the accuracy of the frozen-scattering approximation in this ideal neutral model. It is \emph{not} interpreted as a scaling law for a separately evaluated continuum force. Exact real-energy poles can enter or leave the gap as parameters vary, transferring spectral weight between discrete poles and the continuum; only their sum is smooth and unambiguous. The small remainder nevertheless demonstrates that the leading neutral force is already encoded in the explicit $L$ dependence of the evanescent transmission eigenvalues.

This behavior differs from a length-independent metallic point-contact model, where the zeroth-order short-junction ABS energy has no leading longitudinal length derivative and above-gap energy dependence becomes essential.\cite{BeenakkerCJ2023,Levchenko2025} The graphene result should therefore be phrased as a statement about the \emph{full force approaching the evanescent short-junction law}, rather than as an exact universal partition of the force into bound and continuum pieces.

\FloatBarrier
\subsection{Thermal suppression and Andreev-character diagnostics}

\begin{figure}[!tb]
\centering
\includegraphics[width=\linewidth]{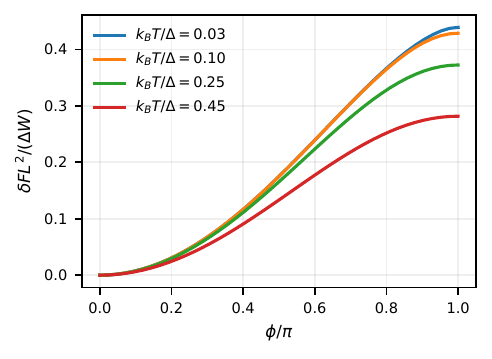}
\caption{Temperature dependence of the neutral total force ($L/\xi=0.30$, $W/L=20$). Normalized force $\delta F\,L^2/(\Delta W)$ versus phase $\phi/\pi$ for four temperatures $\kB T/\Delta=0.03$, $0.10$, $0.25$, and $0.45$. Raising the temperature thermally depopulates the Andreev levels and rounds the response near $\phi=\pi$, lowering the peak force while leaving the small-phase behavior essentially unchanged.}
\label{fig:temperature}
\end{figure}

Figure~\ref{fig:temperature} shows that temperature mainly suppresses and rounds the large-phase response. At $\phi=\pi$, the normalized force decreases from $0.439$ at $\kB T/\Delta=0.03$ to $0.282$ at $\kB T/\Delta=0.45$. The characteristic thermal scale is the smaller of the superconducting gap and the phase-dependent Andreev minigap. At neutrality the minigap closes for the most transparent modes as $\phi\rightarrow\pi$, explaining the enhanced thermal sensitivity near that point.

\begin{figure}[!tb]
\centering
\includegraphics[width=\linewidth]{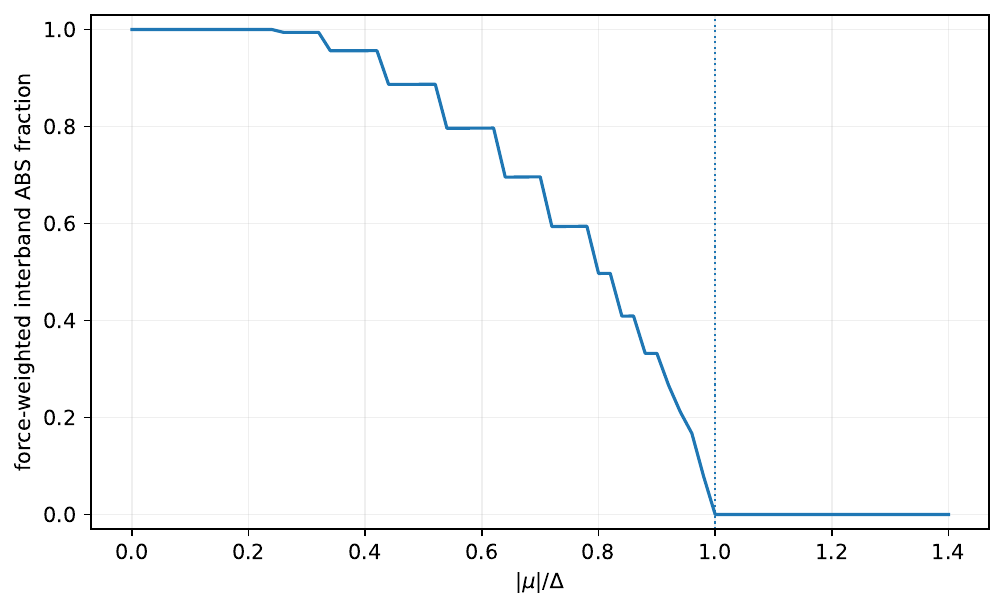}
\caption{Force-weighted interband content of the frozen-scattering short-junction ABS sector. The vertical axis shows the fraction of the absolute short-junction ABS force carried by modes satisfying $E_n^{(0)}>|\mu|$, plotted versus $|\mu|/\Delta$. The dotted line marks $|\mu|/\Delta=1$. This quantity summarizes the crossover within the frozen-scattering ABS sector and is not used as a unique decomposition of the full equilibrium force.}
\label{fig:interbandfrac}
\end{figure}

Figure~\ref{fig:interbandfrac} provides a complementary force-weighted interband diagnostic of the frozen-scattering short-junction ABS sector. Each mode is weighted by the absolute magnitude of its force contribution and classified as interband/specular when $E_n^{(0)}> |\mu|$. The resulting fraction decreases monotonically from nearly unity at charge neutrality to zero near $|\mu|/\Delta=1$, showing that the short-junction ABS response evolves continuously from predominantly interband to predominantly intraband as the junction is doped. This one-dimensional measure gives an intuitive summary of the crossover while remaining an interpretive diagnostic rather than a thermodynamic decomposition of the full force.

\begin{figure}[!tb]
\centering
\includegraphics[width=\linewidth]{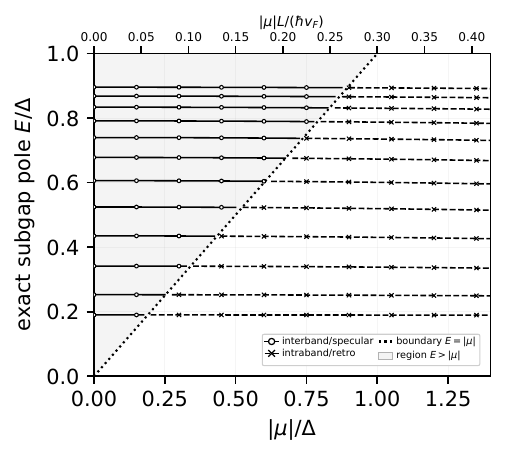}
\caption{Exact-pole Andreev-character diagnostic for the first twelve transverse modes at $L/\xi=0.30$, $W/L=20$, and $\phi=0.85\pi$. Solid/open-circle segments are interband/specular; dashed/cross segments are intraband/retro. The lightly shaded region satisfies $E>|\mu|$, the dotted line is $E=|\mu|$, and the upper axis shows $|\mu|L/(\hbar v_F)$. This exact-pole plot provides a microscopic verification of the crossover summarized more simply in Fig.~\ref{fig:interbandfrac}. No additive partial-force decomposition is implied.}
\label{fig:character}
\end{figure}

Figure~\ref{fig:character} then provides the microscopic refinement of that picture by showing the exact positive subgap poles obtained from Eq.~\eqref{eq:secular} for the first twelve transverse modes. Each pole is classified directly from the Dirac band indices $s_e=\sgn(\mu+E)$ and $s_h=\sgn(\mu-E)$: solid segments with open circles have $s_es_h<0$ and are interband/specular, while dashed segments with crosses have $s_es_h>0$ and are intraband/retro. The diagonal $E=|\mu|$ is therefore the microscopic band-character boundary, not a phenomenological crossover weight. Pole branches cross this line on the scale $|\mu|/\Delta$.

This band-character scale should not be confused with the gate scale of the Fabry-P\'erot oscillations, $|\mu|L/(\hbar v_F)$. For $L/\xi\ll1$, the condition $|\mu|\sim\Delta$ corresponds to $|\mu|L/(\hbar v_F)\sim L/\xi\ll1$. The exact subgap spectrum can therefore become predominantly intraband before the junction reaches the strongly oscillatory many-propagating-mode regime. Because the total equilibrium determinant coherently combines all channels and the above-gap spectrum, Fig.~\ref{fig:character} is used only as a spectral diagnostic; no additive ``retro force'' or ``specular force'' is inferred in the mixed regime.

\section{Experimental scale and detection protocol}

The neutral analytical maximum provides a useful estimate,
\begin{equation}
\delta F_{\max}\simeq0.441\frac{\Delta W}{L^2}.
\label{eq:estimate}
\end{equation}
For aluminum contacts with $\Delta\simeq\SI{0.20}{meV}$, $L=\SI{200}{nm}$, and $W=\SI{4}{\micro m}$, Eq.~\eqref{eq:estimate} gives approximately $\SI{1.4}{fN}$. For a larger-gap contact with $\Delta\simeq\SI{1.0}{meV}$, $L=\SI{100}{nm}$, and $W=\SI{5}{\micro m}$, the estimate is approximately $\SI{35}{fN}$. These values are phase-dependent corrections to much larger static mechanical and electrostatic forces.

A practical detection strategy is therefore lock-in measurement under phase modulation,
\begin{equation*}
\phi(t)=\phi_0+\delta\phi\cos\omega t.
\end{equation*}
The force component at the modulation frequency is
\begin{equation*}
\delta F_\omega=
\left.\frac{\partial F}{\partial\phi}\right|_{\phi_0}\delta\phi
=-\frac{\hbar}{2e}
\left.\frac{\partial I}{\partial L}\right|_{\phi_0}\delta\phi.
\end{equation*}
A suspended graphene membrane or compliant superconducting electrode could in principle convert this longitudinal force into a measured mechanical coordinate.\cite{Bunch2007} The intrinsic force scale can exceed representative thermomechanical force-noise estimates, but a device-specific transduction calculation is required because $F=-\partial_L\Omega$ is conjugate to the longitudinal junction length, whereas many graphene resonators are read out through flexural motion. A realistic experiment must therefore specify how longitudinal stress modifies the measured mode frequency or displacement. Gate-driven sign changes of $\delta F$, electron-hole symmetry, and phase locking provide discriminants against static electrostatic attraction and the electromagnetic Casimir background.

The sharp ideal resonances should be compared with three broadening scales: $\kB T$, the elastic scattering rate $\hbar/\tau_e$, and the gate-inhomogeneity energy. Observing individual resonances requires all three to remain smaller than the longitudinal mode spacing $\hbar v_F/L$. Even when that condition is not satisfied, a smooth gate dependence and a neutral-to-doped change in force amplitude should remain.

\section{Limitations and extensions}

The model is deliberately minimal. First, the superconducting gap is imposed as a step function and is not calculated self-consistently. Inverse proximity effects can modify both the local gap and its length derivative. Second, the contact regions are assumed infinitely doped and the electrostatic profile is abrupt. Recent calculations show that smoothing the graphene Josephson potential profile modifies conductance, critical current, and current-phase skewness,\cite{Rycerz2026} so the narrow resonances predicted here should be regarded as an ideal coherent limit. Third, interactions, strain, pseudomagnetic fields, real magnetic fields, and spin-orbit coupling are neglected. Fourth, the mechanical derivative holds $\mu$ fixed. In a device controlled at fixed gate voltage, motion can also change the capacitance and hence the carrier density, producing an additional electromechanical term that must be separated experimentally.

The present Matsubara formulation can incorporate these effects by replacing Eqs.~\eqref{eq:Se} to \eqref{eq:r} with the appropriate energy-dependent normal-region scattering matrix. Interface barriers can be concatenated with $S_e$ and $S_h$ before evaluating Eq.~\eqref{eq:M}. Disorder can be treated by recursive Green functions or scattering-matrix ensembles. A self-consistent calculation requires differentiating the stationary free energy, including the $L$ dependence of the order parameter and electrostatic profile.

The exact pole solver of Eq.~\eqref{eq:secular} also clarifies why a smooth universal split into ``bound'' and ``continuum'' forces is delicate: as a pole crosses $|E|=\Delta$, its contribution is transferred to the continuum scattering phase, producing compensating threshold terms. A projector-resolved real-energy spectral-shift calculation could further separate interband, intraband, and interference weights, but only their total is a unique equilibrium force in the mixed regime.

\section{Conclusions}

We have formulated and evaluated an energy-dependent BdG scattering determinant for the phase-dependent equilibrium force correction of an ideal ballistic SGS junction. At charge neutrality the complete Matsubara result approaches the analytical evanescent-mode law $\delta F\propto\Delta W/L^2$, with the explicit length dependence of $\tau(q)=\sech^2(qL)$ supplying the leading short-junction mechanical sensitivity. Gate doping produces a qualitatively different regime in which propagating channels generate Fabry-P\'erot resonances and sign changes of $\delta F=F(\phi)-F(0)$.

We have also separated two notions that are easily conflated. Exact real-energy subgap poles change from interband/specular to intraband/retro character on the scale $|\mu|/\Delta$, while strong propagation resonances are controlled by $|\mu|L/(\hbar v_F)$. These scales are parametrically distinct for $L/\xi\ll1$. Finally, we have avoided identifying the difference between the full determinant and the frozen-scattering short-junction result with a pure continuum force; that remainder contains both finite-energy subgap corrections and above-gap phase shifts.

The combination of a neutral evanescent-mode force law, a gate-controlled coherent resonance structure, and a microscopic Andreev-character crossover constitutes a graphene-specific mechanical response complementary to the Josephson current. Experimental observability will depend not only on the intrinsic femtonewton-scale thermodynamic force but also on how the longitudinal junction stress couples to a measurable mechanical mode.

\begin{acknowledgments} 
SS was supported by a KIAS individual Grant (PG099101) at the Korea Institute for Advanced Study. 
\end{acknowledgments}

\appendix

\section{Reduction of the determinant to the universal short-junction spectrum}
\label{app:reduction}

If the normal scattering matrix is energy independent on the scale $\Delta$, time-reversal symmetry permits its polar decomposition in terms of a transmission eigenvalue $\tau$. Evaluating Eq.~\eqref{eq:M} gives
\begin{equation*}
\frac{\det\mcM(\phi)}{\det\mcM(0)}
=\frac{\omega_m^2+\Delta^2[1-\tau\sin^2(\phi/2)]}
{\omega_m^2+\Delta^2}.
\end{equation*}
The Matsubara product therefore has poles at
\begin{equation*}
E=\pm\Delta\sqrt{1-\tau\sin^2\frac{\phi}{2}}.
\end{equation*}
recovering Eq.~\eqref{eq:ABS}. This check also fixes the plus sign in Eq.~\eqref{eq:M} after analytic continuation of the Andreev amplitude.

\section{Derivation of the neutral analytical integral}

At zero temperature, the phase-dependent free energy of the positive BdG levels is
\begin{align*}
\delta\Om_{\mathrm{ABS}}
&=\frac{g}{2}\frac{W}{\pi}\int_0^\infty\dd q
\left[\Delta-E_q(\phi)\right]\\
&=\frac{2\Delta W}{\pi L}J(s),
\end{align*}
where $x=qL$ and
\begin{equation*}
J(s)=\int_0^\infty\dd x
\left[1-\sqrt{1-s^2\sech^2x}\right].
\end{equation*}
Differentiating with respect to $s$ gives
\begin{align*}
\frac{\dd J}{\dd s}
&=s\int_0^\infty\frac{\sech^2x\,\dd x}
{\sqrt{1-s^2\sech^2x}}\\
&=s\int_0^1\frac{\dd t}{\sqrt{1-s^2+s^2t^2}}
=\artanh s,
\end{align*}
where $t=\tanh x$. Since $J(0)=0$,
\begin{equation*}
J(s)=s\artanh s+\frac12\ln(1-s^2).
\end{equation*}
which proves Eqs.~\eqref{eq:neutralOmega} to \eqref{eq:neutralF}.

\section{Analytical transmission derivatives}

For a propagating mode, write
\begin{equation*}
\tau=(1+A\sin^2k_xL)^{-1},\qquad A=q^2/k_x^2.
\end{equation*}
At fixed $q$ and $\mu$,
\begin{equation}
\partial_L\tau=-A k_x\sin(2k_xL)\tau^2.
\label{eq:dtauprop}
\end{equation}
The sign alternates with the Fabry-P\'erot phase. For an evanescent mode,
\begin{equation}
\tau=(1+B\sinh^2\kappa L)^{-1},\qquad B=q^2/\kappa^2,
\end{equation}
so
\begin{equation}
\partial_L\tau=-\frac{q^2}{\kappa}\sinh(2\kappa L)\tau^2<0.
\label{eq:dtauevan}
\end{equation}
The smooth sign of Eq.~\eqref{eq:dtauevan} and the oscillating sign of Eq.~\eqref{eq:dtauprop} explain the main qualitative contrast between the neutral and doped force curves.

\section{Exact real-energy pole diagnostic}
\label{app:poles}

The exact spectral diagnostic used in Fig.~\ref{fig:character} is obtained from the real-energy continuation of the same round-trip matrix used on the Matsubara axis. For $0<E<\Delta$, the Andreev amplitude has unit modulus, $\alpha(E)=\exp[-\ii\arccos(E/\Delta)]$, and the secular determinant is Eq.~\eqref{eq:secular}. The root search uses $2600$ points uniformly spaced in $y=\arccos(E/\Delta)$ rather than a uniform energy grid, which resolves poles close to the gap edge. Candidate minima with $|D_n|<2.5\times10^{-2}$ are refined by bounded minimization to a tolerance $\Delta y=10^{-13}$ and accepted when the refined determinant satisfies $|D_n|<3\times10^{-7}$. Multiplicity is diagnosed from secular-matrix singular values smaller than $2\times10^{-5}$. Doubling the Andreev-angle grid to $5200$ points for representative modes $n=0,3,6,9,11$ and gate values $|\mu|/\Delta=0.2,0.6,1.0,1.3$ shifts the accepted poles by less than $2\times10^{-8}\Delta$.

For each accepted positive pole the band character is determined directly from $s_e=\sgn(\mu+E)$ and $s_h=\sgn(\mu-E)$. No artificial broadening parameter or hyperbolic-tangent crossover weight is introduced. When a pole approaches $E=\Delta$ it may merge with the continuum as $L$, $\phi$, or $\mu$ is varied. A force obtained by differentiating the \emph{bound-state sector alone} can therefore contain threshold terms that are canceled by the continuum spectral phase shift. This is why the main text reports the complete Matsubara force and treats Eq.~\eqref{eq:remainder} only as an approximation error/remainder rather than as an exact continuum force.

\section{Numerical implementation and convergence}

The calculation uses positive Matsubara frequencies and positive transverse indices only, with $g=4$. The transverse cutoff is chosen as
\begin{equation*}
q_{\max}=\frac{qL_{\max}}{L}+|\mu|/(\hbar v_F).
\end{equation*}
with $qL_{\max}=14$ for the main figures. The Matsubara sum is truncated at $\omega_{\max}=80\Delta$ to $100\Delta$. The centered derivative uses $\delta L/L=7.5\times10^{-4}$ and keeps the transverse mode set fixed between $L-\delta L$ and $L+\delta L$.

At $L/\xi=0.50$, $W/L=20$, $\phi=0.9\pi$, and $\kB T/\Delta=0.05$, changing $qL_{\max}$ from $14$ to $16$ changes the normalized neutral force by less than $3\times10^{-11}$ and the result at $\mu/\Delta=6$ by less than $2\times10^{-13}$. Increasing $\omega_{\max}$ from $40\Delta$ to $100\Delta$ changes the displayed values below the quoted precision. The sharp intraband resonance near $\mu L/(\hbar v_F)=2.25$ (at $\mu/\Delta=4.5$), which carries the largest force in Fig.~\ref{fig:gate}, is likewise stable: its normalized peak value is unchanged to five significant figures as $qL_{\max}$ is varied from $12$ to $18$ and $\omega_{\max}$ from $60\Delta$ to $120\Delta$, confirming that the resonance is a converged physical feature rather than a discretization artifact. Electron-hole symmetry is verified numerically, and the neutral determinant is checked against Eq.~\eqref{eq:neutralF}. The exact pole routine locates zeros of Eq.~\eqref{eq:secular} using the same energy-dependent scattering matrices; its energy grid is uniform in $\arccos(E/\Delta)$ to resolve states arbitrarily close to the gap edge.

\end{document}